\documentclass[fleqn]{llncs}
\usepackage{pgajs}

\title{Program Algebra with a Jump-Shift Instruction%
       \thanks{This research has been partly carried out in the
               framework of the  Jacquard-project Symbiosis, which is
               funded by the Netherlands Organisation for Scientific
               Research (NWO).}}
\author{J.A. Bergstra\inst{1}\fnmsep\inst{2}
        \and
        C.A. Middelburg\inst{1}
       }
\institute{Programming Research Group,
           University of Amsterdam, \\
           P.O.~Box~41882, 1009~DB~Amsterdam, the Netherlands \\
           \and
           Department of Philosophy,
           Utrecht University, \\
           P.O.~Box~80126, 3508~TC~Utrecht, the Netherlands \\
           \email{J.A.Bergstra@uva.nl,C.A.Middelburg@uva.nl}
          }

\begin{document}
\maketitle

\begin{abstract}
We study sequential programs that are instruction sequences with
jump-shift instructions in the setting of \PGA\ (ProGram Algebra).
Jump-shift instructions preceding a jump instruction increase the
position to jump to.
The jump-shift instruction is not found in programming practice.
Its merit is that the expressive power of \PGA\ extended with the
jump-shift instruction, is not reduced if the reach of jump instructions
is bounded.
This is used to show that there exists a finite-state execution
mechanism that by making use of a counter can produce each finite-state
thread from some program that is a finite or periodic infinite sequence
of instructions from a finite set.
\\[1.5ex]
{\sl Keywords:}
\sloppy
program algebra, jump-shift instruction,
thread algebra, thread extraction,
execution mechanism.
\\[1.5ex]
{\sl 1998 ACM Computing Classification:}
D.3.1, D.3.3, F.1.1, F.3.2, F.3.3.
\end{abstract}

\section{Introduction}
\label{sect-intro}

In this paper, we study sequential programs that are instruction
sequences with jump-shift instructions.
With that we carry on the line of research with which a start was made
in~\cite{BL02a}.
The object pursued with this line of research is the development of a
theoretical understanding of possible forms of sequential programs,
starting from the simplest form.
The view is taken that sequential programs in the simplest form are
sequences of instructions.
\PGA, an algebra of programs in which programs are looked upon as
sequences of instructions, is taken for the basis of the development
aimed at.
The work presented in this paper is part of an investigation of the
consequences of small differences in the choice of primitives in the
algebra of programs.

In the line of research carried on in this paper, the view is taken that
the behaviours of sequential programs under execution are threads as
considered in basic thread algebra~\cite{BL02a}.%
\footnote
{In~\cite{BL02a}, basic thread algebra is introduced under the name
 basic polarized process algebra.
 Prompted by the development of thread algebra~\cite{BM04c}, which is a
 design on top of it, basic polarized process algebra has been renamed
 to basic thread algebra.
}
The experience gained so far leads us to believe that sequential
programs are nothing but linear representations of threads.

If $n$ jump-shift instructions precede a jump instruction, they increase
the position to jump to by $n$.
This feature of the jump-shift instruction is called jump shifting.
It is a programming feature that is not suggested by existing
programming practice.
Its merit is that the expressive power of \PGA\ extended with the
jump-shift instruction, unlike the expressive power of \PGA, is not
reduced if the reach of jump instructions is bounded.
Therefore, we consider a study of programs that are instruction
sequences with jump-shift instructions relevant to programming.

We believe that interaction with services provided by an execution
environment is inherent in the behaviour of programs under execution.
Intuitively, a counter service provides for jump shifting.
In this paper, we define the meaning of programs with jump-shift
instructions in two different ways.
One way covers all programs with jump-shift instructions.
The other way covers all programs with jump-shift instructions that
contain no other jump instruction than the one whose effect in the
absence of preceding jump-shift instructions is a jump to the position
of the instruction itself.
The latter way corresponds to a finite-state execution mechanism that by
making use of a counter produces the behaviour of a program from that
program.

A thread proceeds by doing steps in a sequential fashion.
A thread may do certain steps only for the sake of getting reply values
returned by some service and that way having itself affected by that
service.
The interaction between behaviours of programs under execution and a
counter service referred to above is an interaction with that purpose.
In~\cite{BM04c}, the use mechanism is introduced to allow for such a
kind of interaction between threads and services.
In this paper, we will use the use mechanism, which has been renamed to
thread-service composition, to have behaviours of programs under
execution affected by services.

A hierarchy of program notations rooted in \PGA\ is introduced
in~\cite{BL02a}.
Included in this hierarchy are very simple program notations which are
close to existing assembly languages up to and including simple program
notations that support structured programming by offering a rendering of
conditional and loop constructs.
However, although they are found in existing assembly programming
practice, indirect jump instructions are not considered.
In~\cite{BM07e}, several kinds of indirect jump instructions are
considered, including a kind by which recursive method calls can easily
be explained.
Moreover, dynamic instruction instantiation, a useful programming
feature that is not suggested by existing programming practice, is
considered in~\cite{BM07f}.

This paper is organized as follows.
First, we review basic thread algebra (Section~\ref{sect-BTA}).
Next, we review \PGA\ and extend it with the jump-shift instruction
(Section~\ref{sect-PGAjs}).
After that, we extend basic thread algebra with thread-service
composition, introduce a state-based approach to describe services, and
give a state-based description of counter services
(Section~\ref{sect-TAtsc}).
Following this, we revisit the meaning of programs with jump-shift
instructions and show that there exists a finite-state execution
mechanism that by making use of a counter can produce each finite-state
thread from a program that is a finite or periodic infinite sequence of
instructions from a finite set (Section~\ref{sect-revisited}).
Finally, we make some concluding remarks (Section~\ref{sect-concl}).

\section{Basic Thread Algebra}
\label{sect-BTA}

In this section, we review \BTA\ (Basic Thread Algebra), a form of
process algebra which is tailored to the description of the behaviour of
deterministic sequential programs under execution.
The behaviours concerned are called \emph{threads}.

In \BTA, it is assumed that there is a fixed but arbitrary finite set
$\BAct$ of \emph{basic actions} with $\Tau \not\in \BAct$.
We write $\BActTau$ for $\BAct \union \set{\Tau}$.
The members of $\BActTau$ are referred to as \emph{actions}.

The intuition is that each basic action performed by a thread is taken
as a command to be processed by a service provided by the execution
environment of the thread.
The processing of a command may involve a change of state of the service
concerned.
At completion of the processing of the command, the service produces a
reply value.
This reply is either $\True$ or $\False$ and is returned to the thread
concerned.

Although \BTA\ is one-sorted, we make this sort explicit.
The reason for this is that we will extend \BTA\ with an additional sort
in Section~\ref{sect-TAtsc}.

The algebraic theory \BTA\ has one sort: the sort $\Thr$ of
\emph{threads}.
To build terms of sort $\Thr$, \BTA\ has the following constants and
operators:
\begin{iteml}
\item
the \emph{deadlock} constant $\const{\DeadEnd}{\Thr}$;
\item
the \emph{termination} constant $\const{\Stop}{\Thr}$;
\item
for each $a \in \BActTau$, the binary \emph{postconditional composition}
operator $\funct{\pcc{\ph}{a}{\ph}}{\linebreak[2]\Thr \x \Thr}{\Thr}$.
\end{iteml}
Terms of sort $\Thr$ are built as usual (see e.g.~\cite{ST99a,Wir90a}).
Throughout the paper, we assume that there are infinitely many variables
of sort $\Thr$, including $x,y,z$.

We use infix notation for postconditional composition.
We introduce \emph{action prefixing} as an abbreviation: $a \bapf p$,
where $p$ is a term of sort $\Thr$, abbreviates $\pcc{p}{a}{p}$.

Let $p$ and $q$ be closed terms of sort $\Thr$ and $a \in \BActTau$.
Then $\pcc{p}{a}{q}$ will perform action $a$, and after that proceed as
$p$ if the processing of $a$ leads to the reply $\True$ (called a
positive reply), and proceed as $q$ if the processing of $a$ leads to
the reply $\False$ (called a negative reply).
The action $\Tau$ plays a special role.
It is a concrete internal action: performing $\Tau$ will never lead to a
state change and always lead to a positive reply, but notwithstanding
all that its presence matters.

\BTA\ has only one axiom.
This axiom is given in Table~\ref{axioms-BTA}.%
\begin{table}[!tb]
\caption{Axiom of \BTA}
\label{axioms-BTA}
\begin{eqntbl}
\begin{axcol}
\pcc{x}{\Tau}{y} = \pcc{x}{\Tau}{x}                      & \axiom{T1}
\end{axcol}
\end{eqntbl}
\end{table}
Using the abbreviation introduced above, axiom T1 can be written as
follows: $\pcc{x}{\Tau}{y} = \Tau \bapf x$.

Each closed \BTA\ term of sort $\Thr$ denotes a finite thread, i.e.\ a
thread of which the length of the sequences of actions that it can
perform is bounded.
Guarded recursive specifications give rise to infinite threads.

A \emph{recursive specification} over \BTA\ is a set of recursion
equations $\set{X = t_X \where X \in V}$ where $V$ is a set of variables
of sort $\Thr$ and each $t_X$ is a \BTA\ term of sort $\Thr$ that
contains only variables from $V$.
Let $E$ be a recursive specification over \BTA.
Then we write $\vars(E)$ for the set of all variables that occur on the
left-hand side of an equation in $E$.
A \emph{solution} of a recursive specification $E$ is a set of threads
(in some model of \BTA) $\set{T_X \where X \in \vars(E)}$ such that
the equations of $E$ hold if, for all $X \in \vars(E)$, $X$ stands for
$T_X$.

Let $t$ be a \BTA\ term of sort $\Thr$ containing a variable $X$ of sort
$\Thr$.
Then an occurrence of $X$ in $t$ is \emph{guarded} if $t$ has a subterm
of the form $\pcc{t'}{a}{t''}$ containing this occurrence of $X$.
Let $E$ be a recursive specification over \BTA.
Then $E$ is a \emph{guarded recursive specification} if, in each
equation $X = t_X \in E$, all occurrences of variables in $t_X$ are
guarded or $t_X$ can be rewritten to such a term using the equations in
$E$ from left to right.
We are only interested in models of \BTA\ in which guarded recursive
specifications have unique solutions, such as the projective limit model
of \BTA\ presented in~\cite{BB03a}.
A thread that is the solution of a finite guarded recursive
specification over \BTA\ is called a \emph{finite-state} thread.

We extend \BTA\ with guarded recursion by adding constants for solutions
of guarded recursive specifications and axioms concerning these
additional constants.
For each guarded recursive specification $E$ and each $X \in \vars(E)$,
we add a constant of sort $\Thr$ standing for the unique solution of $E$
for $X$ to the constants of \BTA.
The constant standing for the unique solution of $E$ for $X$ is denoted
by $\rec{X}{E}$.
Moreover, we add the axioms for guarded recursion given in
Table~\ref{axioms-rec} to \BTA.%
\begin{table}[!tb]
\caption{Axioms for guarded recursion}
\label{axioms-rec}
\begin{eqntbl}
\begin{saxcol}
\rec{X}{E} = \rec{t_X}{E} & \mif X \!=\! t_X \in E       & \axiom{RDP}
\\
E \Implies X = \rec{X}{E} & \mif X \in \vars(E)          & \axiom{RSP}
\end{saxcol}
\end{eqntbl}
\end{table}
In this table, we write $\rec{t_X}{E}$ for $t_X$ with, for all
$Y \in \vars(E)$, all occurrences of $Y$ in $t_X$ replaced by
$\rec{Y}{E}$.
$X$, $t_X$ and $E$ stand for an arbitrary variable of sort $\Thr$, an
arbitrary \BTA\ term of sort $\Thr$ and an arbitrary guarded recursive
specification over \BTA, respectively.
Side conditions are added to restrict the variables, terms and guarded
recursive specifications for which $X$, $t_X$ and $E$ stand.
The equations $\rec{X}{E} = \rec{t_X}{E}$ for a fixed $E$ express that
the constants $\rec{X}{E}$ make up a solution of $E$.
The conditional equations $E \Implies X = \rec{X}{E}$ express that this
solution is the only one.

We will write \BTA+\REC\ for \BTA\ extended with the constants for
solutions of guarded recursive specifications and axioms RDP and RSP.
We will often write $X$ for $\rec{X}{E}$ if $E$ is clear from the
context.
It should be borne in mind that, in such cases, we use $X$ as a
constant.
We will also use the following abbreviation: $a^\omega$, where
$a \in \BActTau$, abbreviates $\rec{X}{\set{X = a \bapf X}}$.

In~\cite{BM05c}, we show that the threads considered in \BTA+\REC\ can
be viewed as processes that are definable over ACP~\cite{Fok00}.

Closed terms of sort $\Thr$ from the language of \BTA+\REC\ that denote
the same infinite thread cannot always be proved equal by means of the
axioms of \BTA+\REC.
We introduce the approximation induction principle to remedy this.
The approximation induction principle, \AIP\ in short, is based on the
view that two threads are identical if their approximations up to any
finite depth are identical.
The approximation up to depth $n$ of a thread is obtained by cutting it
off after performing a sequence of actions of length $n$.

\AIP\ is the infinitary conditional equation given in
Table~\ref{axioms-AIP}.%
\begin{table}[!tb]
\caption{Approximation induction principle}
\label{axioms-AIP}
\begin{eqntbl}
\begin{axcol}
\AND{n \geq 0} \proj{n}{x} = \proj{n}{y} \Implies x = y   & \axiom{AIP}
\end{axcol}
\end{eqntbl}
\end{table}
Here, following~\cite{BL02a}, approximation of depth $n$ is phrased in
terms of a unary \emph{projection} operator
$\funct{\pi_n}{\Thr}{\Thr}$.
The axioms for the projection operators are given in
Table~\ref{axioms-pin}.%
\begin{table}[!tb]
\caption{Axioms for projection operators}
\label{axioms-pin}
\begin{eqntbl}
\begin{axcol}
\proj{0}{x} = \DeadEnd                                   & \axiom{P0} \\
\proj{n+1}{\Stop} = \Stop                                & \axiom{P1} \\
\proj{n+1}{\DeadEnd} = \DeadEnd                          & \axiom{P2} \\
\proj{n+1}{\pcc{x}{a}{y}} =
                       \pcc{\proj{n}{x}}{a}{\proj{n}{y}} & \axiom{P3}
\end{axcol}
\end{eqntbl}
\end{table}
In this table, $a$ stands for an arbitrary member of $\BActTau$.

We will write \BTA+\REC+\AIP\ for \BTA+\REC\ extended with the
projection operators and the axioms from Tables~\ref{axioms-AIP}
and~\ref{axioms-pin}.

A \emph{linear recursive specification} over \BTA\ is a guarded
recursive specification $E = \set{X = t_X \where X \in V}$, where each
$t_X$ is a term of the form $\DeadEnd$, $\Stop$ or $\pcc{Y}{a}{Z}$ with
$Y,Z \in V$.
For each closed term $p$ of sort $\Thr$ from the language of \BTA+\REC,
there exist a linear recursive specification $E$ and a variable
$X \in \vars(E)$ such that $p = \rec{X}{E}$ is derivable from the axioms
of \BTA+\REC.

Below, the interpretations of the constants and operators of \BTA+\REC\
in models of \BTA+\REC\ are denoted by the constants and operators
themselves.
Let $\cA$ be some model of \BTA+\REC, and let $p$ be an element from the
domain of $\cA$.
Then the set of \emph{states} or \emph{residual threads} of $p$,
written $\Res(p)$, is inductively defined as follows:
\begin{iteml}
\item
$p \in \Res(p)$;
\item
if $\pcc{q}{a}{r} \in \Res(p)$, then $q \in \Res(p)$ and
$r \in \Res(p)$.
\end{iteml}
We are only interested in models of \BTA+\REC\ in which
$\card(\Res(\rec{X}{E})) \leq \card(E)$ for all finite linear recursive
specifications $E$, such as the projective limit model of \BTA\ presented
in~\cite{BB03a}.

\section{Program Algebra and the Jump-Shift Instruction}
\label{sect-PGAjs}

In this section, we first review \PGA\ (ProGram Algebra) and then extend
it with the jump-shift instruction, resulting in \PGAjs.
\PGA\ is an algebra of sequential programs based on the idea that
sequential programs are in essence sequences of instructions.
\PGA\ provides a program notation for finite-state threads.
The jump-shift instruction is not found in programming practice: if one
or more jump-shift instructions precede a jump instruction, then each of
those jump-shift instructions increases the position to jump to by one.

\subsection{PGA}
\label{subsect-PGA}

In \PGA, it is assumed that there is a fixed but arbitrary finite set
$\BInstr$ of \emph{basic instructions}.
\PGA\ has the following \emph{primitive instructions}:
\begin{iteml}
\item
for each $a \in \BInstr$, a \emph{plain basic instruction} $a$;
\item
for each $a \in \BInstr$, a \emph{positive test instruction} $\ptst{a}$;
\item
for each $a \in \BInstr$, a \emph{negative test instruction} $\ntst{a}$;
\item
for each $l \in \Nat$, a \emph{forward jump instruction} $\fjmp{l}$;
\item
a \emph{termination instruction} $\halt$.
\end{iteml}
We write $\PJInstr$ for the set of all forward jump instructions and
$\PInstr_\sPGA$ for the set of all primitive instructions of \PGA.

The intuition is that the execution of a basic instruction $a$ may
modify a state and produces $\True$ or $\False$ at its completion.
In the case of a positive test instruction $\ptst{a}$, basic instruction
$a$ is executed and execution proceeds with the next primitive
instruction if $\True$ is produced and otherwise the next primitive
instruction is skipped and execution proceeds with the primitive
instruction following the skipped one.
In the case where $\True$ is produced and there is not at least one
subsequent primitive instruction and in the case where $\False$ is
produced and there are not at least two subsequent primitive
instructions, deadlock occurs.
In the case of a negative test instruction $\ntst{a}$, the role of the
value produced is reversed.
In the case of a plain basic instruction $a$, the value produced is
disregarded: execution always proceeds as if $\True$ is produced.
The effect of a forward jump instruction $\fjmp{l}$ is that execution
proceeds with the $l$-th next instruction of the program concerned.
If $l$ equals $0$ or the $l$-th next instruction does not exist, then
$\fjmp{l}$ results in deadlock.
The effect of the termination instruction $\halt$ is that execution
terminates.

\PGA\ has the following constants and operators:
\begin{iteml}
\item
for each $u \in \PInstr_\sPGA$, an \emph{instruction} constant $u$\,;
\item
the binary \emph{concatenation} operator $\ph \conc \ph$\,;
\item
the unary \emph{repetition} operator $\ph\rep$\,.
\end{iteml}
Terms are built as usual.
Throughout the paper, we assume that there are infinitely many
variables, including $x,y,z$.

We use infix notation for concatenation and postfix notation for
repetition.
We also use the notation $P^n$.
For each \PGA\ term $P$ and $n > 0$, the term $P^n$ is defined by
induction on $n$ as follows: $P^1 = P$ and $P^{n+1} = P \conc P^n$.

Closed \PGA\ terms are considered to denote programs.
The intuition is that a program is in essence a non-empty, finite or
periodic infinite sequence of primitive instructions.%
\footnote
{A periodic infinite sequence is an infinite sequence with only finitely
 many subsequences.}
These sequences are called \emph{single pass instruction sequences}
because \PGA\ has been designed to enable single pass execution of
instruction sequences: each instruction can be dropped after it has been
executed.
Programs are considered to be equal if they represent the same single
pass instruction sequence.
The axioms for instruction sequence equivalence are given in
Table~\ref{axioms-PGA}.%
\begin{table}[!t]
\caption{Axioms of \PGA}
\label{axioms-PGA}
\begin{eqntbl}
\begin{axcol}
(x \conc y) \conc z = x \conc (y \conc z)              & \axiom{PGA1} \\
(x^n)\rep = x\rep                                      & \axiom{PGA2} \\
x\rep \conc y = x\rep                                  & \axiom{PGA3} \\
(x \conc y)\rep = x \conc (y \conc x)\rep              & \axiom{PGA4}
\end{axcol}
\end{eqntbl}
\end{table}
In this table, $n$ stands for an arbitrary natural number greater than
$0$.
The \emph{unfolding} equation $x\rep = x \conc x\rep$ is
derivable.
Each closed \PGA\ term is derivably equal to a term in
\emph{canonical form}, i.e.\ a term of the form $P$ or $P \conc Q\rep$,
where $P$ and $Q$ are closed \PGA\ terms that do not contain the
repetition operator.

Each closed \PGA\ term is considered to denote a program of which the
behaviour is a finite-state thread, taking the set $\BInstr$ of basic
instructions for the set $\BAct$ of actions.
The \emph{thread extraction} operation $\extr{\ph}$ assigns a thread to
each program.
The thread extraction operation is defined by the equations given in
Table~\ref{axioms-thread-extr} (for $a \in \BInstr$, $l \in \Nat$ and
$u \in \PInstr_\sPGA$)%
\begin{table}[!t]
\caption{Defining equations for thread extraction operation}
\label{axioms-thread-extr}
\begin{eqntbl}
\begin{eqncol}
\extr{a} = a \bapf \DeadEnd \\
\extr{a \conc x} = a \bapf \extr{x} \\
\extr{\ptst{a}} = a \bapf \DeadEnd \\
\extr{\ptst{a} \conc x} =
\pcc{\extr{x}}{a}{\extr{\fjmp{2} \conc x}} \\
\extr{\ntst{a}} = a \bapf \DeadEnd \\
\extr{\ntst{a} \conc x} =
\pcc{\extr{\fjmp{2} \conc x}}{a}{\extr{x}}
\end{eqncol}
\qquad
\begin{eqncol}
\extr{\fjmp{l}} = \DeadEnd \\
\extr{\fjmp{0} \conc x} = \DeadEnd \\
\extr{\fjmp{1} \conc x} = \extr{x} \\
\extr{\fjmp{l+2} \conc u} = \DeadEnd \\
\extr{\fjmp{l+2} \conc u \conc x} = \extr{\fjmp{l+1} \conc x} \\
\extr{\halt} = \Stop \\
\extr{\halt \conc x} = \Stop
\end{eqncol}
\end{eqntbl}
\end{table}
and the rule given in Table~\ref{rule-thread-extr}.%
\begin{table}[!t]
\caption{Rule for cyclic jump chains}
\label{rule-thread-extr}
\begin{eqntbl}
\begin{eqncol}
x \scongr \fjmp{0} \conc y \Implies \extr{x} = \DeadEnd
\end{eqncol}
\end{eqntbl}
\end{table}
This rule is expressed in terms of the \emph{structural congruence}
predicate $\ph \scongr \ph$, which is defined by the formulas given in
Table~\ref{axioms-scongr} (for $n,m,l \in \Nat$ and
$u_1,\ldots,u_n,v_1,\ldots,v_{m+1} \in \PInstr_\sPGA$).%
\begin{table}[!t]
\caption{Defining formulas for structural congruence predicate}
\label{axioms-scongr}
\begin{eqntbl}
\begin{eqncol}
\fjmp{n+1} \conc u_1 \conc \ldots \conc u_n \conc \fjmp{0}
\scongr
\fjmp{0} \conc u_1 \conc \ldots \conc u_n \conc \fjmp{0}
\\
\fjmp{n+1} \conc u_1 \conc \ldots \conc u_n \conc \fjmp{m}
\scongr
\fjmp{m+n+1} \conc u_1 \conc \ldots \conc u_n \conc \fjmp{m}
\\
(\fjmp{n+l+1} \conc u_1 \conc \ldots \conc u_n)\rep \scongr
(\fjmp{l} \conc u_1 \conc \ldots \conc u_n)\rep
\\
\fjmp{m+n+l+2} \conc u_1 \conc \ldots \conc u_n \conc
(v_1 \conc \ldots \conc v_{m+1})\rep \scongr {} \\ \hfill
\fjmp{n+l+1} \conc u_1 \conc \ldots \conc u_n \conc
(v_1 \conc \ldots \conc v_{m+1})\rep
\\
x \scongr x
\\
x_1 \scongr y_1 \And x_2 \scongr y_2 \Implies
x_1 \conc x_2 \scongr y_1 \conc y_2 \And
{x_1}\rep \scongr {y_1}\rep
\end{eqncol}
\end{eqntbl}
\end{table}

The equations given in Table~\ref{axioms-thread-extr} do not cover the
case where there is a cyclic chain of forward jumps.
Programs are structural congruent if they are the same after removing
all chains of forward jumps in favour of single jumps.
Because a cyclic chain of forward jumps corresponds to $\fjmp{0}$,
the rule from Table~\ref{rule-thread-extr} can be read as follows:
if $x$ starts with a cyclic chain of forward jumps, then $\extr{x}$
equals $\DeadEnd$.
It is easy to see that the thread extraction operation assigns the same
thread to structurally congruent programs.
Therefore, the rule from Table~\ref{rule-thread-extr} can be replaced by
the following generalization:
$x \scongr y  \Implies \extr{x} = \extr{y}$.

Let $E$ be a finite guarded recursive specification over \BTA, and let
$P_X$ be a closed \PGA\ term for each $X \in \vars(E)$.
Let $E'$ be the set of equations that results from replacing in $E$ all
occurrences of $X$ by $\extr{P_X}$ for each $X \in \vars(E)$.
If $E'$ can be obtained by applications of axioms PGA1--PGA4, the
defining equations for the thread extraction operation and the rule for
cyclic jump chains, then $\extr{P_X}$ is the solution of $E$ for $X$.
Such a finite guarded recursive specification can always be found.
Thus, the behaviour of each closed \PGA\ term, is a thread that is
definable by a finite guarded recursive specification over \BTA.
Moreover, each finite guarded recursive specification over \BTA\ can be
translated to a closed \PGA\ term of which the behaviour is the solution
of the finite guarded recursive specification concerned
(cf.\ Section~4 of~\cite{PZ06a}).

Closed \PGA\ terms are loosely called \PGA\ \emph{programs}.
\PGA\ programs in which the repetition operator do not occur
are called \emph{finite} \PGA\ programs.

\subsection{The Jump-Shift Instruction}
\label{subsect-jump-shift-instr}

We extend \PGA\ with the jump-shift instruction, resulting in \PGAjs.

In \PGAjs, like in \PGA, it is assumed that there is a fixed but
arbitrary finite set $\BInstr$ of \emph{basic instructions}.
\PGAjs\ has the \emph{primitive instructions} of \PGA\ and in addition:
\begin{iteml}
\item
a \emph{jump-shift instruction} $\shift$.
\end{iteml}
We write $\PInstr_\sPGAjs$ for the set of all primitive instructions of
\PGAjs.

If one or more jump-shift instructions precede a jump instruction, then
each of those jump-shift instructions increases the position to jump to
by one.
If one or more jump-shift instructions precede an instruction different
from a jump instruction, then those jump-shift instructions have no
effect.

\PGAjs\ has the following constants and operators:
\begin{iteml}
\item
for each $u \in \PInstr_\sPGAjs$, an \emph{instruction} constant $u$\,;
\item
the binary \emph{concatenation} operator $\ph \conc \ph$\,;
\item
the unary \emph{repetition} operator $\ph\rep$\,.
\end{iteml}

The axioms of \PGAjs\ are the axioms of \PGA\ (Table~\ref{axioms-PGA})
and in addition the axioms for the jump-shift instruction given in
Table~\ref{axioms-jump-shift}.%
\begin{table}[!t]
\caption{Additional axioms for the jump-shift instruction}
\label{axioms-jump-shift}
\begin{eqntbl}
\begin{axcol}
\shift \conc \fjmp{l} = \fjmp{l+1}                     & \axiom{JSI1} \\
\shift \conc u = u                                     & \axiom{JSI2} \\
\shift\rep = \fjmp{0}\rep                              & \axiom{JSI3}
\end{axcol}
\end{eqntbl}
\end{table}
In this table, $u$ stands for an arbitrary primitive instruction from
$\PInstr_\sPGA \diff \PJInstr$.

The thread extraction operation of \PGAjs\ is defined by the same
equations and rule as the thread extraction operation of \PGA\
(Tables~\ref{axioms-thread-extr} and~\ref{rule-thread-extr}), on the
understanding that $u$ still stands for an arbitrary primitive
instruction from $\PInstr_\sPGA$, and in addition the equation given in
Table~\ref{axiom-thread-extr-add}.%
\begin{table}[!t]
\caption{Additional defining equation for thread extraction operation}
\label{axiom-thread-extr-add}
\begin{eqntbl}
\begin{eqncol}
\extr{x} = \extr{x \conc \fjmp{0}}
\end{eqncol}
\end{eqntbl}
\end{table}
The structural congruence predicate of \PGAjs\ is defined by the same
formulas as the structural congruence predicate of \PGA\
(Table~\ref{axioms-scongr}), on the understanding that $u_1$, \ldots,
$u_n$, $v_1$, \ldots, $v_{m+1}$ still stand for arbitrary primitive
instructions from $\PInstr_\sPGA$.

The additional defining equation $\extr{x} = \extr{x \conc \fjmp{0}}$
for the thread extraction operation expresses that a missing termination
instructions leads to deadlock.
For all \PGA\ programs $P$, the equation
$\extr{P} = \extr{P \conc \fjmp{0}}$ is derivable from the axioms of
\PGA\ and the defining equations for the thread extraction operation of
\PGA.
For all \PGAjs\ programs $P$, the equation
$\extr{\fjmp{l+2} \conc \shift \conc P} = \extr{\fjmp{l+2} \conc P}$ is
derivable from the axioms of \PGAjs\ and the defining equations of the
thread extraction operation of \PGAjs.

Obviously, the set of all \PGA\ programs is a proper subset of the set
of all \PGAjs\ programs.
Moreover, the thread extraction operation of \PGA\ is the restriction of
the thread extraction operation of \PGAjs\ to the set of all \PGA\
programs.
Therefore, we do not distinguish the two thread extraction operations
syntactically.

Below, we consider \PGAjs\ programs that contain no other jump
instruction than $\fjmp{0}$.
We will refer to these programs as \PGAjsz\ programs.

An interesting point of \PGAjsz\ programs is that they make use of a
finite set of primitive instructions.
It happens that, although \PGA\ programs make use of an infinite set of
primitive instructions, \PGA\ programs do not offer more expressive
power than \PGAjsz\ programs.
\begin{theorem}
\label{theorem-expr-power}
Each \PGA\ program $P$ can be transformed to a \PGAjsz\ program $P'$
such that $\extr{P} = \extr{P'}$.
\end{theorem}
\begin{proof}
Let $P$ be a \PGA\ program, and
let $P'$ be $P$ with, for all $l > 0$, all occurrences of $\fjmp{l}$ in
$P$ replaced by $\shift^l \conc \fjmp{0}$.
Clearly, $P'$ is a \PGAjsz\ program.
It is easily proved by induction on $l$ that, for each $l > 0$, the
equation $\fjmp{l} = \shift^l \conc \fjmp{0}$ is derivable from the
axioms of \PGAjs.
From this it follows immediately that the equation $P = P'$ is derivable
from the axioms of \PGAjs.
Consequently, $\extr{P} = \extr{P'}$.
\qed
\end{proof}
As a corollary of Theorem~\ref{theorem-expr-power} and the
expressiveness results for \PGA\ in~\cite{PZ06a}, we have that
$\extr{\ph}$ can produce each finite-state thread from some \PGAjsz\
program.
\begin{corollary}
\label{corollary-expr-power}
For each finite-state thread $p$, there exists a \PGAjsz\ program $P$
such that $\extr{P} = p$.
\end{corollary}
This means that each finite-state thread can be produce from a program
that is a finite or periodic infinite sequence of instructions from a
finite set.

\subsection{On Single Pass Execution of Instruction Sequences}
\label{subsect-single-pass}

The primitive instructions of \PGA\ have been designed to enable single
pass execution of instruction sequences.
Thread extraction defined in accordance with the idea of single pass
execution of instruction sequences should ideally only involve equations
of the form $\extr{u \conc x} = p$ where $\extr{x}$ is the only
expression of the form $\extr{P}$ that may occur in $p$.
In this section, thread extraction has not been defined in accordance
with the idea of single pass execution of instruction sequences.
The equations
$\extr{\ptst{a} \conc x} = \pcc{\extr{x}}{a}{\extr{\fjmp{2} \conc x}}$,
$\extr{\ntst{a} \conc x} = \pcc{\extr{\fjmp{2} \conc x}}{a}{\extr{x}}$,
and $\extr{\fjmp{l+2} \conc u \conc x} = \extr{\fjmp{l+1} \conc x}$ are
not of the right form.
In Section~\ref{sect-revisited}, we define an alternative thread
extraction operation for \PGAjsz\ programs, which is better in
accordance with the idea of single pass execution of instruction
sequences.
By that thread extraction operation, each \PGAjsz\ program is assigned a
thread that becomes the behaviour that it exhibits on execution by
interaction with a counter service.
In Section~\ref{sect-TAtsc}, we introduce thread-service composition,
which allows for the intended interaction.

\section{Services and Interaction of Threads with Services}
\label{sect-TAtsc}

In this section, we first extend \BTA\ with thread-service composition,
next introduce a state-based approach to describe services, and then use
this approach to give a description of counter services.
In the current paper, we will only use thread-service composition to
have program behaviours affected by some service.

\subsection{Thread-Service Composition}
\label{subsect-TAtsc}

A thread may perform certain actions only for the sake of getting reply
values returned by services and that way having itself affected by
services.
We introduce thread-service composition to allow for threads to be
affected in this way.
Thread-service composition is introduced under the name \emph{use}
in~\cite{BM04c}.

It is assumed that there is a fixed but arbitrary finite set $\Foci$ of
\emph{foci} and a fixed but arbitrary finite set $\Meth$ of
\emph{methods}.
Each focus plays the role of a name of a service provided by the
execution environment that can be requested to process a command.
Each method plays the role of a command proper.
For the set $\BAct$ of actions, we take the set
$\set{f.m \where f \in \Foci, m \in \Meth}$.
Performing an action $f.m$ is taken as making a request to the
service named $f$ to process command $m$.

We introduce yet another sort: the sort $\Serv$ of \emph{services}.
However, we will not introduce constants and operators to build terms
of this sort.
$\Serv$ is considered to stand for the set of all services.
We identify services with functions
$\funct{H}{\neseqof{\Meth}}{\set{\True,\False,\Blocked}}$
that satisfy the following condition:
\begin{ldispl}
\Forall{\alpha \in \neseqof{\Meth},m \in \Meth}
 {(H(\alpha) = \Blocked \Implies H(\alpha \concat \seq{m}) =
   \Blocked)}\;.
\end{ldispl}
Given a service $H$ and a method $m \in \Meth$,
the \emph{derived service} of $H$ after processing $m$,
written $\derive{m}H$, is defined by
$\derive{m}H(\alpha) = H(\seq{m} \concat \alpha)$.

A service $H$ can be understood as follows:
\begin{iteml}
\item
if $H(\seq{m}) = \True$, then the request to process $m$ is accepted by
the service, the reply is positive, and the service proceeds as
$\derive{m}H$;
\item
if $H(\seq{m}) = \False$, then the request to process $m$ is accepted by
the service, the reply is negative, and the service proceeds as
$\derive{m}H$;
\item
if $H(\seq{m}) = \Blocked$, then the request to process $m$ is rejected
by the service.
\end{iteml}

For each $f \in \Foci$, we introduce the binary
\emph{thread-service composition} operator
$\funct{\use{\ph}{f}{\ph}}{\Thr \x \Serv}{\Thr}$.
Intuitively, $\use{p}{f}{H}$ is the thread that results from processing
all actions performed by thread $p$ that are of the form $f.m$ by
service $H$.
When an action of the form $f.m$ performed by thread $p$ is processed by
service $H$, it is turned into $\Tau$ and postconditional composition is
removed in favour of action prefixing on the basis of the reply value
produced.

The axioms for the thread-service composition operators are given in
Table~\ref{axioms-tsc}.%
\begin{table}[!t]
\caption{Axioms for thread-service composition}
\label{axioms-tsc}
\begin{eqntbl}
\begin{saxcol}
\use{\Stop}{f}{H} = \Stop                            & & \axiom{TSC1} \\
\use{\DeadEnd}{f}{H} = \DeadEnd                      & & \axiom{TSC2} \\
\use{\Tau \bapf x}{f}{H} =
                          \Tau \bapf (\use{x}{f}{H}) & & \axiom{TSC3} \\
\use{(\pcc{x}{g.m}{y})}{f}{H} =
\pcc{(\use{x}{f}{H})}{g.m}{(\use{y}{f}{H})}
 & \mif f \neq g                                       & \axiom{TSC4} \\
\use{(\pcc{x}{f.m}{y})}{f}{H} =
\Tau \bapf (\use{x}{f}{\derive{m}H})
 & \mif H(\seq{m}) = \True                             & \axiom{TSC5} \\
\use{(\pcc{x}{f.m}{y})}{f}{H} =
\Tau \bapf (\use{y}{f}{\derive{m}H})
 & \mif H(\seq{m}) = \False                            & \axiom{TSC6} \\
\use{(\pcc{x}{f.m}{y})}{f}{H} = \DeadEnd
 & \mif H(\seq{m}) = \Blocked                          & \axiom{TSC7}
\end{saxcol}
\end{eqntbl}
\end{table}
In this table, $f$ and $g$ stand for an arbitrary foci from $\Foci$ and
$m$ stands for an arbitrary method from $\Meth$.
Axioms TSC3 and TSC4 express that the action $\Tau$ and actions of
the form $g.m$, where $f \neq g$, are not processed.
Axioms TSC5 and TSC6 express that a thread is affected by a service
as described above when an action of the form $f.m$ performed by the
thread is processed by the service.
Axiom TSC7 expresses that deadlock takes place when an action to be
processed is not accepted.

Let $T$ stand for either \BTA, \BTA+\REC\ or \BTA+\REC+\AIP.
Then we will write $T$+\TSC\ for $T$, taking the set
$\set{f.m \where f \in \Foci, m \in \Meth}$ for $\BAct$, extended with
the thread-service composition operators and the axioms from
Table~\ref{axioms-tsc}.

The action $\Tau$ is an internal action whose presence matters.
To conceal its presence in the case where it does not matter after all,
we also introduce the unary \emph{abstraction} operator
$\funct{\abstr}{\Thr}{\Thr}$.

The axioms for the abstraction operator are given in
Table~\ref{axioms-abstr}.%
\begin{table}[!t]
\caption{Axioms for abstraction}
\label{axioms-abstr}
\begin{eqntbl}
\begin{axcol}
\abstr(\Stop) = \Stop                                    & \axiom{TT1} \\
\abstr(\DeadEnd) = \DeadEnd                              & \axiom{TT2} \\
\abstr(\Tau \bapf x) = \abstr(x)                         & \axiom{TT3} \\
\abstr(\pcc{x}{a}{y}) = \pcc{\abstr(x)}{a}{\abstr(y)}    & \axiom{TT4}
\end{axcol}
\end{eqntbl}
\end{table}
In this table, $a$ stands for an arbitrary basic action from $\BAct$.

Abstraction can for instance be appropriate in the case where $\Tau$
arises from turning actions of an auxiliary nature into $\Tau$ on
thread-service composition.
Examples of this case will occur in Section~\ref{sect-revisited}.
Unlike the use mechanism introduced in~\cite{BM04c}, the use mechanism
introduced in~\cite{BP02a} incorporates abstraction.

Let $T$ stand for either \BTA, \BTA+\REC, \BTA+\REC+\AIP, \BTA+\TSC,
BTA+\REC+\TSC\ or \BTA+\REC+\AIP+\TSC.
Then we will write $T$+\ABSTR\ for $T$ extended with the abstraction
operator and the axioms from Table~\ref{axioms-abstr}.

The equation $\abstr(\Tau^\omega) = \DeadEnd$ is derivable from the
axioms of \BTA+\REC+\linebreak[2]\AIP+\ABSTR.

\subsection{State-Based Description of Services}
\label{subsect-service-descr}

We introduce a state-based approach to describe families of services
which will be used in Section~\ref{subsect-counter}.
The approach is similar to the approach to describe state machines
introduced in~\cite{BP02a}.

In this approach, a family of services is described by
\begin{itemize}
\item
a set of states $S$;
\item
an effect function $\funct{\eff}{\Meth \x S}{S}$;
\item
a yield function
$\funct{\yld}{\Meth \x S}{\set{\True,\False,\Blocked}}$;
\end{itemize}
satisfying the following condition:
\begin{ldispl}
\Exists{s \in S}
 {\Forall{m \in \Meth}
   {{} \\ \quad
    (\yld(m,s) = \Blocked \And
     \Forall{s' \in S}
      {(\yld(m,s') = \Blocked \Implies \eff(m,s') = s)})}}\;.
\end{ldispl}
The set $S$ contains the states in which the services may be, and the
functions $\eff$ and $\yld$ give, for each method $m$ and state $s$, the
state and reply, respectively, that result from processing $m$ in state
$s$.

We define, for each $s \in S$, a cumulative effect function
$\funct{\ceff_s}{\seqof{\Meth}}{S}$ in terms of $s$ and $\eff$ as follows:
\begin{ldispl}
\ceff_s(\emptyseq) = s\;,
\\
\ceff_s(\alpha \concat \seq{m}) = \eff(m,\ceff_s(\alpha))\;.
\end{ldispl}
We define, for each $s \in S$, a service $H_s$ in terms of $\ceff_s$ and
$\yld$ as follows:
\begin{ldispl}
H(\alpha \concat \seq{m})  = \yld(m,\ceff_s(\alpha))\;.
\end{ldispl}
$H_s$ is called the service with \emph{initial state} $s$ described by
$S$, $\eff$ and $\yld$.
We say that $\set{H_s \where s \in S}$ is the \emph{family of services}
described by $S$, $\eff$ and $\yld$.

The condition that is imposed on $S$, $\eff$ and $\yld$ imply that, for
each $s \in S$, $H_s$ is a service indeed.
It is worth mentioning that $\derive{m} H_s = H_{\eff(m,s)}$ and
$H(\seq{m}) = \yld(m,s)$.

\subsection{Counter Services}
\label{subsect-counter}

We give a state-based description of a very simple family of services
that constitute a counter.
This counter will be used in Section~\ref{sect-revisited} to describe
the behaviour of programs in \PGAjs.

The counter services accept the following methods:
\begin{itemize}
\item
a \emph{counter reset method} $\clrc$;
\item
a \emph{counter increment method} $\incc$;
\item
a \emph{counter decrement method} $\decc$;
\item
a \emph{counter is-zero method} $\iszc$.
\end{itemize}
We write $\Meth_\cnt$ for the set
$\set{\clrc,\incc,\decc,\iszc}$.
It is assumed that $\Meth_\cnt \subseteq \Meth$.

The methods accepted by counter services can be explained as follows:
\begin{itemize}
\item
$\clrc$\,:
the content of the counter is set to zero and the reply is $\True$;
\item
$\incc$\,:
the content of the counter is incremented by one and the reply is
$\True$;
\item
$\decc$\,:
if the content of the counter is greater than zero, then the content of
the counter is decremented by one and the reply is $\True$; otherwise,
nothing changes and the reply is $\False$;
\item
$\iszc$\,:
if the content of the counter equals zero, then nothing changes and the
reply is $\True$; otherwise, nothing changes and the reply is $\False$.
\end{itemize}

Let $s \in \Nat$.
Then we write $\Cnt_s$ for the service with initial state $s$ described
by $S = \Nat \union \set{\undef}$, where $\undef \not\in \Nat$, and the
functions $\eff$ and $\yld$ defined as follows ($k \in \Nat$):
\begin{ldispl}
\begin{gceqns}
\eff(\clrc,k) = 0\;,
\\
\eff(\incc,k) = k + 1\;,
\\
\eff(\decc,0) = 0\;,
\\
\eff(\decc,k + 1) = k\;,
\\
\eff(\iszc,k) = k\;,
\\ {} \\
\eff(m,k)     = \undef   & \mif m \not\in \Meth_\cnt\;,
\\
\eff(m,\undef)= \undef\;,
\end{gceqns}
\quad\;
\begin{gceqns}
\yld(\clrc,k) = \True\;,
\\
\yld(\incc,k) = \True\;,
\\
\yld(\decc,0) = \False\;,
\\
\yld(\decc,k + 1) = \True\;,
\\
\yld(\iszc,0) = \True\;,
\\
\yld(\iszc,k + 1) = \False\;,
\\
\yld(m,k)      = \Blocked & \mif m \not\in \Meth_\cnt\;,
\\
\yld(m,\undef) = \Blocked\;.
\end{gceqns}
\end{ldispl}
We write $\Cnt_\mathrm{init}$ for $\Cnt_0$.

\section{\PGAjsz\ Programs Revisited}
\label{sect-revisited}

In this section, we define an alternative thread extraction operation
for \PGAjsz\ programs, which is in accordance with the idea of single
pass execution of instruction sequences.
By that thread extraction operation, each \PGAjsz\ program is assigned a
thread that becomes the behaviour that it exhibits on execution by
interaction with a counter service.
We also introduce a notion of an execution mechanism.
The alternative thread extraction operation induces a finite-state
execution mechanism that by making use of a counter can produce each
finite-state thread from some \PGAjsz\ program.

\subsection{Alternative Semantics for \PGAjsz\ Programs}
\label{subsects-alternative}

When defining the alternative thread extraction operation for \PGAjsz\
programs, it is assumed that there is a fixed but arbitrary finite set
$\Foci$ of foci with $\cnt \in \Foci$ and a fixed but arbitrary finite
set $\Meth$ of methods.
Besides, the set
$\set{f.m \where f \in \Foci \diff \set{\cnt}, m \in \Meth}$
is taken as the set $\BInstr$ of basic instructions.
Thereby no real restriction is imposed on the set $\BInstr$: in the case
where the cardinality of $\Foci$ equals~$2$, all basic instructions have
the same focus and the set $\Meth$ of methods can be looked upon as the
set $\BInstr$ of basic instructions.

The alternative thread extraction operation $\extrg{\ph}$ for \PGAjsz\
programs is defined by the equations given in
Table~\ref{axioms-thread-extr-alt} (for $a \in \BInstr$, $l \in \Nat$,
$u \in (\PInstr_\sPGAjs \diff \PJInstr) \union \set{\fjmp{0}}$).%
\begin{table}[!t]
\caption{Defining equations for thread extraction operation}
\label{axioms-thread-extr-alt}
\begin{eqntbl}
\begin{seqncol}
\extrg{x} = \extrg{x \conc \fjmp{0}} \\
\extrg{a \conc x} = \cnt.\clrc \bapf (a \bapf \extrg{x}) \\
\extrg{\ptst{a} \conc x} =
\cnt.\clrc \bapf (\pcc{\extrg{x}}{a}{(\cnt.\incc \bapf \extrsk{x})}) \\
\extrg{\ntst{a} \conc x} =
\cnt.\clrc \bapf (\pcc{(\cnt.\incc \bapf \extrsk{x})}{a}{\extrg{x}}) \\
\extrg{\shift \conc x} = \cnt.\incc \bapf \extrg{x} \\
\extrg{\fjmp{0} \conc x} = \pcc{\DeadEnd}{\cnt.\iszc}{\extrsk{x}} \\
\extrg{\halt \conc x} = \Stop
\eqnsep
\extrsk{\shift \conc x} = \extrsk{x} \\
\extrsk{u \conc x} =
\cnt.\decc \bapf (\pcc{\extrg{u \conc x}}{\cnt.\iszc}{\extrsk{x}})
 & \mif u \neq \shift
\end{seqncol}
\end{eqntbl}
\end{table}
The thread assigned to a program by this thread extraction operation is
not the behaviour that the program exhibits on execution.
That behaviour arises from interaction of this thread with a counter
service.

The following theorem states rigorously that, for any \PGAjsz\ program,
the behaviour under execution coincides with the alternative behaviour
under execution on interaction with a counter when abstracted from
$\Tau$.
\begin{theorem}
\label{theorem-behaviour}
For all \PGAjsz\ programs $P$,
$\extr{P} = \abstr(\use{\extrg{P}}{\cnt}{\Cnt_\mathrm{init}})$.
\end{theorem}
\begin{proof}
Strictly speaking, we prove this theorem in the algebraic theory
obtained by:
(i)~combining \PGAjs\ with \BTA+\REC+\AIP+\TSC+\ABSTR, resulting in a
theory with three sorts: a sort $\Prog$ of programs, a sort $\Thr$ of
threads, and a sort $\Serv$ of services;
(ii)~extending the result by taking $\extr{\ph}$ and $\extrg{\ph}$ for
additional operators from sort $\Prog$ to sort $\Thr$ and taking the
semantic equations and rule defining thread extraction and alternative
thread extraction for additional axioms.
We write $\TProg$ for the set of all closed terms of sort $\Prog$ from
the language of the resulting theory and $\TThr$ for the set of all
closed terms of sort $\Thr$ from the language of the resulting theory.
Moreover, we write $\TProgz$ for the set of all closed terms from
$\TProg$ that contain no other jump instructions than $\fjmp{0}$.

Let
\begin{ldispl}
T =
\set{\extr{P},\extr{\shift^{i+1} \conc P},
     \extr{\shift^{i+1} \conc \fjmp{0} \conc P}
     \where i \in \Nat \And P \in \TProgz}\;, \\
T' =
\set{\abstr(\use{\extrg{P}}{\cnt}{\Cnt_i}),
     \abstr(\use{\extrsk{P}}{\cnt}{\Cnt_{i+1}})
     \where i \in \Nat \And P \in \TProgz}\;,
\end{ldispl}
and let $\funct{\beta}{T}{T'}$ be the bijection defined by
\begin{ldispl}
\beta(\extr{P}) =
\abstr(\use{\extrg{P}}{\cnt}{\Cnt_\mathrm{init}})\;, \\
\beta(\extr{\shift^{i+1} \conc P}) =
\abstr(\use{\extrg{P}}{\cnt}{\Cnt_{i+1}})\;, \\
\beta(\extr{\shift^{i+1} \conc \fjmp{0} \conc P}) =
\abstr(\use{\extrsk{P}}{\cnt}{\Cnt_{i+1}})\;.
\end{ldispl}
For each $p' \in \TThr$, write $\beta^*(p')$ for $p'$ with, for all
$p \in T$, all occurrences of $p$ in $p'$ replaced by $\beta(p)$.
Then, it is straightforward to prove that there exists a set $E$
consisting of one derivable equation $p = p'$ for each $p \in T$ such
that, for all equations $p = p'$ in $E$:
\begin{iteml}
\item
the equation $\beta(p) = \beta^*(p')$ is also derivable;
\item
if $p' \in T$, then $p'$ can always be rewritten to a $p'' \not\in T$
using the equations in $E$ from left to right.
\end{iteml}
Because
$\beta(\extr{P}) = \abstr(\use{\extrg{P}}{\cnt}{\Cnt_\mathrm{init}})$,
this means that, for all $P \in \TProgz$, $\extr{P}$ and
$\abstr(\use{\extrg{P}}{\cnt}{\Cnt_\mathrm{init}})$ are solutions of the
same guarded recursive specification.
Because guarded recursive specifications have unique solutions, it
follows immediately that, for all $P \in \TProgz$,
$\extr{P} = \abstr(\use{\extrg{P}}{\cnt}{\Cnt_\mathrm{init}})$.
\qed
\end{proof}
As a corollary of Theorem~\ref{theorem-behaviour} and
Corollary~\ref{corollary-expr-power}, we have that $\extrg{\ph}$ by
making use of a counter can produce each finite-state thread from some
\PGAjsz\ program.
\begin{corollary}
\label{corollary-behaviour}
For each finite-state thread $p$, there exists a \PGAjsz\ program $P$
such that $\abstr(\use{\extrg{P}}{\cnt}{\Cnt_\mathrm{init}}) = p$.
\end{corollary}

\subsection{On Finite-State Execution Mechanisms}
\label{subsect-execution}

Below, we introduce a notion of an execution mechanism.
The intuition is that, for a function that assigns a finite-state
behaviour to each member of some set of instruction sequences, an
execution mechanism is a deterministic behaviour that can produce the
behaviour assigned to each of these instruction sequences from the
instruction sequence concerned by going through the instructions in the
sequence one by one.
We believe that there do not exist execution mechanisms that can deal
with sequences of instructions from an infinite set.
Therefore, we restrict ourselves to finite instruction sets.

Let $\PInstr$ be a finite set, let $\cP$ be a set of non-empty finite or
periodic infinite sequences over $\PInstr$, and let $\extr{\ph}$ be a
function that assigns a finite-state thread to each member of $\cP$.
Assume that $\pgs \in \Foci$, that $\hdeq{:}u \in \Meth$ for all
$u \in \PInstr$, that $\drop \in \Meth$, and that basic actions of the
form $\pgs.m$ do not occur in $\extr{P}$ for all $P \in \cP$.
Moreover, for each $P \in \cP$, let $\PGS_P$ be the service with initial
state $P$ described by
$S = \cP \union \set{\empseq} \union \set{\undef}$, where
$\undef \not\in \cP \union \set{\empseq}$,%
\footnote{We write $\empseq$ for the empty sequence.}
and the functions $\eff$ and $\yld$ defined as follows
($u,u' \in \PInstr$, $P \in \cP$, $Q \in \cP \union \set{\empseq}$):
\begin{ldispl}
\begin{gceqns}
\eff(\hdeq{:}u,Q) = Q\;,
\\ {} \\ {} \\ {} \\ {} \\
\eff(\drop,\empseq) = \empseq\;,
\\
\eff(\drop,u) = \empseq\;,
\\
\eff(\drop,u \conc P) = P\;,
\\
\eff(m,Q) = \undef   & \mif m \not\in \Meth_\pgs\;,
\\
\eff(m,\undef) = \undef\;,
\end{gceqns}
\;\;
\begin{gceqns}
\yld(\hdeq{:}u,\empseq) = \False\;,
\\
\yld(\hdeq{:}u,u) = \True\;,
\\
\yld(\hdeq{:}u,u \conc P) = \True\;,
\\
\yld(\hdeq{:}u,u') = \False         & \mif u \neq u'\;,
\\
\yld(\hdeq{:}u,u' \conc P) = \False & \mif u \neq u'\;,
\\
\yld(\drop,\empseq) = \False\;,
\\
\yld(\drop,u) = \True\;,
\\
\yld(\drop,u \conc P) = \True\;,
\\
\yld(m,Q) = \Blocked & \mif m \not\in \Meth_\pgs\;,
\\
\yld(m,\undef) = \Blocked\;.
\end{gceqns}
\end{ldispl}
Then an \emph{execution mechanism} for $\extr{\ph}$ is a thread $p$ such
that $\abstr(\use{p}{\pgs}{\PGS_P}) = \extr{P}$ for all $P \in \cP$.
An execution mechanism is called a \emph{finite-state} execution
mechanism if it is a finite-state thread.

In order to execute an instruction sequence $P$, an execution mechanism
makes use of the service $\PGS_P$ to go through that the instructions in
that sequence one by one.
The methods accepted by this service can be explained as follows:
\begin{iteml}
\item
$\hdeq{:}u$\,:
if there is an instruction sequence left and its first instruction is
$u$, then nothing changes and the reply is $\True$; otherwise, nothing
changes and the reply is $\False$;
\item
$\drop$\,:
if there is an instruction sequence left, then its first instruction is
dropped and the reply is $\True$; otherwise, nothing changes and the
reply is $\False$.
\end{iteml}
Notice that the service does not have to hold an infinite object: there
exists an adequate finite representation for each finite or periodic
infinite sequence of instructions.

It is easy to see that there exists a finite-state execution mechanism
for the thread extraction operation $\extrg{\ph}$ for \PGAjsz\ programs.
From this and Corollary~\ref{corollary-behaviour}, it follows
immediately that there exists a finite-state execution mechanism that by
making use of a counter can produce each finite-state thread from some
program that is a finite or periodic infinite sequence of instructions
from a finite set.

We also have that there does not exist a finite-state execution
mechanism that by itself can produce each finite-state thread from a
program that is a finite or periodic infinite sequence of instructions
from a finite set.
\begin{theorem}
\label{theorem-non-existence}
Let $\PInstr$ be a finite set, let $\cP$ be a set of non-empty finite or
periodic infinite sequences over $\PInstr$, and let $\extr{\ph}$ be a
function that assigns a finite-state thread to each member of $\cP$.
Assume that, for each finite-state thread $p$, there exists a
$P \in \cP$ such that $\extr{P} = p$.
Then there does not exist a finite-state execution mechanism for
$\extr{\ph}$.
\end{theorem}
\begin{proof}
Suppose that there exists a finite-state execution mechanism, say
$p_\exec$.
Let $n$ be the number of states of $p_\exec$.
Consider the thread $T_0$ defined by the guarded recursive specification
consisting of the following equations:
\begin{ldispl}
\begin{aceqns}
T_i          & = & \pcc{T_{i+1}}{a}{T'_{i+1,0}}
 & \mathrm{for}\, i \in [0,n]\;,
\\
T_{n+1}      & = & \Stop\;,
\\
T'_{i+1,i'}  & = & b \bapf T'_{i+1,i'+1}
 & \mathrm{for}\, i \in [0,n], i' \in [0,i]\;,
\\
T'_{i+1,i+1} & = & c \bapf T'_{i+1,0}\;.
\end{aceqns}
\end{ldispl}
Let $P$ be a member of $\cP$ from which $p_\exec$ can produce $T_0$.
Notice that $T_0$ performs $a$ at least once and at most $n+1$ times
after each other.
Suppose that $T_0$ has performed $a$ for the $j$th time when the reply
$\False$ is returned, while at that stage $p_\exec$ has gone through the
first $k_j$ instructions of $P$.
Moreover, write $P_j$ for what is left of $P$ after its first $k_j$
instructions have been dropped.
Then $p_\exec$ still has to produce $T'_{j,0}$ from $P_j$.
For each $j \in [1,n+1]$, a $k_j$ as above can be found.
Let $j_0$ be the unique $j \in [1,n+1]$ such that $k_{j'} \leq k_j$ for
all $j' \in [1,n+1]$.
Regardless the number of times $T_0$ has performed $a$ when the reply
$\False$ is returned, $p_\exec$ must eventually have dropped the first
$k_{j_0}$ instructions of $P$.
For each of the $n+1$ possible values of $j$, $p_\exec$ must be in a
different state when $P_{j_0}$ is left, because the thread that
$p_\exec$ still has to produce is different.
However, this is impossible with $n$ states.
\qed
\end{proof}
In the light of Theorem~\ref{theorem-non-existence},
Corollary~\ref{corollary-behaviour} can be considered a positive result:
a finite-state execution mechanism that makes use of a counter is
sufficient.
However, this result is reached at the expense of an extremely
inefficient way of representing jumps.
We do not see how to improve on the linear representation of jumps.
With a logarithmic representation, for instance, we expect that a
counter will not do.

Theorem~\ref{theorem-non-existence} is actually a generalization of
Theorem~4 from~\cite{BP06a} adapted to the current setting.

The hierarchy of program notations rooted in program algebra introduced
in~\cite{BL02a} includes a program notation, called \PGLS, that supports
structured programming by offering a rendering of conditional and loop
constructs instead of (unstructured) jump instructions.
Like \PGAjsz, \PGLS\ has a finite set of primitive instructions.
Like for \smash{\PGAjsz}\ programs, there exists a finite-state
execution mechanism that by making use of a counter can produce the
behaviour of each \PGLS\ program.
However, \PGLS\ programs offer less expressive power than \PGA\
programs (see Section~9 of~\cite{BL02a}).
Therefore, \PGLS\ is unsuited to show that there exists a finite-state
execution mechanism that by making use of a counter can produce each
finite-state thread from some program that is a finite or periodic
infinite sequence of instructions from a finite set.

\section{Conclusions}
\label{sect-concl}

We have studied sequential programs that are instruction sequences with
jump-shift instructions.
We have defined the meaning of the programs concerned in two different
ways which both involve the extraction of threads.
One way covers only programs with jump-shift instructions that contain
no other jump instruction than the one whose effect in the absence of
preceding jump-shift instructions is a jump to the position of the
instruction itself.
We have among other things shown that the extraction of threads involved
in that way corresponds to a finite-state execution mechanism that by
making use of a counter can produce each finite-state thread from some
program that is a finite or periodic infinite sequence of instructions
from a finite set.

In the course of this work, we got convinced that a general format for
the defining equations of thread extraction operations can be devised
that yields thread extraction operations corresponding to execution
mechanisms that can produce each finite-state thread from some program.
One of the options for future work is to investigate this matter.

\bibliographystyle{plain}
\bibliography{TA}


\end{document}